\documentstyle[12pt,epsf]{article}
\def\be{\begin{eqnarray}}
\def\ee{\end{eqnarray}}
\def\eps{\epsilon}
\def\k{\bf k}
\def\kp{{\bf k}^{\prime}}
\def\up{\uparrow}
\def\down{\downarrow}
\def\sig{\bf \sigma}
\def \sumk{\sum\limits_{\k \sig}}
\begin{document}

\begin{flushright}
MRI-PHY/96-39
\end{flushright}
\begin{center}
\Large{{\bf Order Parameter Symmetry in Doped YBCO Systems}}
\end{center}
\begin{center}
\large{\bf {P. K. Mohanty\footnote{Email: peekay@mri.ernet.in}}}
\end{center}
\begin{center}
Mehta Research Institute, Chhatnag Road, Jhusi, Allahabad, 221506, INDIA. 

\vspace{5mm}
\large{\bf{A. Taraphder \footnote{Email: arghya@mri.ernet.in }}}
\end{center}
\begin{center}
Department of Physics \& Meteorology, Indian Institute of Technology,
Kharagpur, 721302, INDIA.
\end{center}
\begin{center}
 and   
\end{center}
\begin{center}
Mehta Research Institute,Chhatnag Road, Jhusi,  Allahabad 221506 India. 
\end{center}

\begin{abstract}
Intense experimental and theoretical efforts have recently been brought
to bear on the determination of the symmetry of superconducting order 
parameter in high T$_c$ Cuprates as this would shed light on the nature
of microscopic interactions that are responsible for superconductivity in
these unusual systems. 
We use a two band model that reproduces the Fermi surface seen in ARPES 
experiments of Shen et al.,\cite{shen} and study the symmetry of order 
parameter in the YBCO system.  The model incorporates the effects of 
tunnelling in the c-direction between the planes and chains 
of these 
Cuprates. It is observed that a suitable choice of the phase of the pair
wave-function in the planes and chains lead to both s- and d-wave like features. 
We calculate the detailed phase diagram of this system, observe the
shift in the position of the node(s) of the order parameter(s) on the 
Fermi surface with doping and explain the seemingly contradictory 
experimental observations about the presence of both s- and/or d-wave 
symmetry\cite{leggett} in these systems.

\end{abstract}
\baselineskip=18pt

    Quite a large number of experiments backed by careful calculations
have been brought to bear upon the resolution of the seemingly
irreconcilable and contradictory observations about the nature of symmetry
of the superconducting order parameter (OP) in the Cuprate
superconductors\cite{leggett}. An understanding of this symmetry, which is
still unresolved, will shed considerable light on the nature of the pairing
mechanism(s) in these unusual systems. 

    Broadly, there are several classes of experiments that probe the
symmetry of OP in the cuprates: i) Transport and thermodynamic
measurements\cite{moler},  notably the temeperature
dependence of the London penetration depth ($\lambda$)\cite{hardy}
that show a linear temperature dependence of $\lambda$, indicating that at
low energies the density of superconducting states (SDOS) varies linearly
with energy in pure samples, a signature of gapless excitations. ii) Angle
resolved photoemission experiments\cite{campuz} that show a node
(with a small offset) along the $(\pi,\pi)$ direction
of the square Fermi surface (FS)of YBa$_2$Cu$_3$O$_7$. iii) Josephson
measurements: magnetic 
field dependence of dc SQUID\cite{wollman,tsuei} predict a
$\pi$ phase shift consistent with d-wave scenario. Conversely, c-axis
Josephson tunneling\cite{sun1} between twinned and untwinned 
YBa$_2$Cu$_3$O$_7$ and
Pb junction finds a Josephson current, albeit small, indicative of a
symmetric OP. Recent atomic forced microscopy measurements of Kitazawa
and his group\cite{kita} found evidence of gap everywhere on the FS in Bi-2212 
materials.  Critical current between a hexagonal 
grain and its surroundings is found to be non-zero\cite{chaudhari} and
it scales with the number of sides engaged, consistent with the presence of 
a conventional s-wave component. iv) Sensitivity of superconducting $T_c$
with elastic scattering is found to be too weak (follows, instead, the
Ioffe-Regel criterion, $k{_f}l=1$, where $l$ is the mean-free path of the
electrons) to conform to the behaviour of a pure d-wave symmetry\cite{sun2}.  

While there is a general concensus that there are nodes on the FS in all
the hole-doped Cuprates, there is, however, no consensus about its origin.
Models based on spin fluctuation\cite{moriya} predict a
$d_{x^{2}-y^{2}}$ symmetry for the OP, while an interlayer tunneling-enhanced 
conventional superconductivity scenario\cite{chakravarty} envisages 
extremely anisotropic s-wave gap function. It is also useful to note that
the symmetry of YBCO system is orthorhombic and not tetragonal, so an
admixture of s- and d-components is inescapable\cite{annet}, and the
nomenclature is based on the dominance of one component over the other. 

In view of this conflicting scenario it is useful to look for models that
incorporate i) the essence of the band structure of the system concerned
and ii) reproduce much of the experimentally observed features.
Comberscot and Leyronas\cite{comber} have tried to incorporate the fact
that in YBCO systems, the existence of both chains and planes can
have a strong bearing on the nature of pairing wave function that occurs.
We use a similar model and show that it is indeed possible to reproduce
many of the experimentally observed features of the OP without explicitly
considering a d-wave pairing function and features consistent with both d-
and s-wave are obtainable with a suitable choice of the relative phase of two
isotropic pairing functions residing mainly on the plane and chain 
derived bands. 

We consider the Hamiltonian: $H= H_{0} + H_{I}$ where, 

\be
H_{0} = \sumk [ 
\eps_{\k} c^{\dag}_{\k \sig} c_{\k \sig}
+\eps^{\prime}_{k_y} d^{\dag}_{k_y \sig} d_{k_y \sig} + t
(d^{\dag}_{k_y \sig}  
c_{\k \sig} + h.c.)] 
\ee
    
\noindent    $\eps_{\k} $ and $\eps^{\prime}_{k_y} $ are energy 
dispersions, and $ c_{\k \sig},  d_{\k \sig} $ are
destruction operators for electrons in the plane and the chain respectively. 
The off-diagonal term $t$ is small ($t\approx 50\,\, meV$)\cite{yu},
momentum independent and introduces 
band interaction. This implies that in $YBCO$ system, an electron is never
in a true eigenstate of planes or chains, rather, it has components on both
the bands. This term shows up in the band-crossing region leading to
anticrossing features with two sheets of the FS as
seen in $YBCO$. Thus moving along a given 
sheet of the FS, an electron changes its character from being in the plane
band to being in the chain band. This feature has its bearing on the
relative sign of the OP as we discuss below. 
   
In order to fit our band structure with a realistic one obtained from the
ARPES data\cite{campuz}  for the $YBCO$ system, we take five nearest 
neighbour hopping
parameters in the plane, in both $a$ and $b$ direction and two nearest
neighbours in the chain. The dispersion relation in the plane is now 
\be
\eps_{\k} & = & -\mu_p -2t_1 {Cos(k_x ) +Cos(k_y)} +4 t_2 Cos(k_x) Cos(k_y)\cr
                  &-&  2t_3{Cos(2k_x)  + Cos(2k_y)}
     -  4t_4[~Cos(2k_x)Cos(k_y)\cr
	&+&Cos(k_x)Cos(2k_y)~] 
                +  4 t_5 Cos(2k_x)Cos(2k_y)  
\ee
\noindent  and in the chain  
\be
 \eps^{\prime}_{k_y} = -\mu_c -2 h_1 Cos(k_y) +2 h_2 Cos(2k_y)
\ee

The resulting FS, shown in Fig. 1, very
closely resembles  the observed FS\cite{campuz} and the position of the
van Hove singularities (about $10 meV$ away from the FS\cite{gofron}) seen in
$YBa_{2}Cu_{3}O_{7}$, with the following choice of the hopping parameters ,\\ 
$[\mu_{p}, t_{1},  ...., t_{5}] =[-290, 300, 100, 25, 25, 20] meV $ and 
$[\mu_{c}, h_{1} , h_{2}] = [-740, 530, 135]meV.$

The non-interacting Hamiltonian $H_{0}$ can be easily diagonalised by the
introduction of new quasiparticle operators $\alpha_{\bf k}$ and
$\beta_{\bf k}$ defined by

\be
\pmatrix{c_{\k \sig}\cr d_{\k \sig} }~~=~~\pmatrix{
\frac{t}{\sqrt{t^{2}+(\eps^+-\eps)^{2}}}&
\frac{\eps^{-}-\eps^{\prime}}{\sqrt{t^{2}+(\eps^{-}-\eps^{\prime})^{2}}} \cr
\frac{\eps^{+}-\eps}{\sqrt{t^{2}+(\eps^+-\eps)^{2}}}&
\frac{t}{\sqrt{t^{2}+(\eps^{-}-\eps^{\prime})^{2}}}  }
\pmatrix{\alpha_{\k \sig}\cr \beta_{\k \sig} },
\ee
The quasiparticles corresponding to $\alpha_{\k \sig}^{\dag}
(\beta_{\k \sig}^{\dag})$ live on the upper (lower) FS given by 
$\eps^{+} =0( \eps^{-} =0)$, where  $\eps^{\pm}~~=~~\frac{1}{2}(\eps+\eps^{\prime})\pm
\sqrt{(\eps   -\eps^{\prime})^{2}+4t^{2} }$ denote the dispersions of the
two quasiparticle bands and the ${\bf k}$-dependences have been omitted for
notational convenience. Fig. 1 shows the corresponding FS in the first
quadrant of the Brillouin zone (BZ) and Fig. 2, the partial and total density of
states (DOS) with the van Hove singularities predominantly located about
$10 meV$ below the FS\cite{gofron}.   

Superconductivity occurs through the interaction part of the
Hamiltonian\cite{suhl} 
\be
H_{I}& = &
-g \sum_{\k \kp} c^{\dag}_{\kp \up} c^{\dag}_{-\kp \down}
c_{-\kp \down} c_{\k \up}
-g^{\prime} \sum_{\k \kp} d^{\dag}_{\kp \up} d^{\dag}_{-\kp \down}
d_{-\kp \down} d_{\k \up} \cr & &
+K \sum_{\k \kp}\left( d^{\dag}_{kp \up} d^{\dag}_{-\kp \down}
c_{-\kp \down} c_{\k \up} +h.c. \right)
\ee

\noindent where $g$ and $g^{\prime}$ are the pairing interactions (attractive,
non-retarded) in the plane and chain respectively; $K$ is the measure of
pairing interaction between plane and chain and is assumed repulsive. It
is this repulsive interaction that forces the quasiparticles in planes and
chains to stay away in real space forcing $\sum_{\bf k}\langle c_{{\bf
k}\sigma}d_{-{\bf k}-\sigma}\rangle\simeq 0$ which requires the
pairing wave function to have nearly equal regions of positive and
negative signs in the BZ.    

A simple BCS mean-field theory gives the following Hamiltonian, 
\be
H^{mf}_{I}~=~
\sum_{\k}\left [ \Delta c^{\dag}_{\k \up} c^{\dag}_{-\k \down} +
\Delta^{\prime}  d^{\dag}_{\k \up} d^{\dag}_{-\k \down} + h.c.\right]
+\Delta^{*} \frac{K\Delta^{\prime}+g^{\prime}\Delta}{gg^{\prime}-K^{2}}
+\Delta^{\prime*}\frac{K\Delta+g\Delta^{\prime}}{gg^{\prime}-K^{2}}
\ee
where $\Delta, \Delta^\prime$ are defined by, 
$$\Delta =-g \sum_{\k}\langle c_{-\k \down} c_{\k \up}\rangle
+K \sum_{\k}\langle d_{-\k \down} d_{\k \up}\rangle$$
$$\Delta^{\prime} =K \sum_{\k}\langle c_{-\k \down} c_{\k \up}\rangle
+g^{\prime}\sum_{\k}\langle d_{-\k \down} d_{\k \up}\rangle$$

Transforming to the quasi-particle operators, $ H^{mf}_{I}$ can be
re-written as 
\be
H^{mf}_{I}~=~
\sum_{\k}\Delta_{\k}  \alpha^{\dag}_{\k \up} \alpha^{\dag}_{-\k \down} +
\sum_{\k}\Delta^{\prime}_{\k}  \beta^{\dag}_{\k \up} \beta^{\dag}_{-\k
\down} + h.c. )+ const.\\ 
\ee

\noindent Here, $\Delta_{\k} =\frac{\Delta t^{2}+\Delta^{\prime}(\eps^{+}-\eps)^{2}} 
{(\eps^{+}-\eps)^{2}+t^{2}}$   and  
$\Delta^{\prime}_{\k} =\frac{\Delta^{\prime} t^{2}+\Delta(
\eps^{+}-\eps)^{2}}{(\eps^{+}-\eps)^{2}+t^{2}}$.
The two bands are completely decoupled above leading to an
identification of $\Delta$ and $\Delta^{\prime}$ as the gap functions in
the plane and the chain respectively.  On the FS $\eps \eps^{\prime}=t^2$ and
one has a single OP given by  $\Delta_{\k}=\Delta^{\prime}_{\k}=
\frac{\Delta^{\prime} \eps +\Delta \eps^{\prime}}{\eps+\eps^{\prime}}.$
Since $\eps,  \eps^{\prime}$ have the same sign on the FS, the only way
$\Delta_{\bf k}$ changes sign (and consequently have a node) along
the way on the FS is by having $\Delta$ and $\Delta^{\prime}$ out of phase. 
This is, of course, ensured by the repulsive interaction $K$ between the
two bands as discussed above.  

Diagonalization of the full Hamiltonian $H_{0} + H^{mf}_{I}$ is achieved
using the usual Nambu representation $\psi^{\dagger}_{\bf
k}=\left(c^{\dagger}_{\k\up} \,c_{-\k\down}\, d^{\dagger}_{\k\up}\,
d_{-\k\down}\right)$ and writing the mean-field Hamiltonian as
$\psi^{\dagger}_{\k} {\cal M} \psi_{\k}$, where $\cal M$ is the $4\times
4$ mean-field Hamiltonian matrix. The four eigenvalues are obtained from

\be
 2{ E^{\pm}_{\k}}^{2}=\eta +\eta^{\prime} \pm \sqrt {  (\eta-\eta^{\prime})^{2}+
4(\tau^{2}+\delta ^{2}) }
\ee
\noindent where, 
$ \eta= \eps^{2} + \Delta^{2} +t^{2},~~
\eta^{\prime}= {\eps^{\prime}}^{2} +{\Delta^{\prime}}^{2}+t^{2}, ~~
\tau =t(\eps+\eps^{\prime})$ and $\delta=t(\Delta-\Delta^{\prime}) $

On minimization of the mean-field free energy with respect to 
$\Delta^{*}$ and $\Delta^{\prime*}$ the gap equations 
are obtained as (with $ \beta$ as the inverse temperature).

\be
\frac{K\Delta+g\Delta^{\prime}}{gg^{\prime}-K^{2}}~=~
\sum_{\k} \left( \frac{\partial E^{+}_{\k}}{\partial
\Delta^{\prime}}tanh\frac{\beta E^{+}_{\k}}{2} + 
\frac{\partial E^{-}_{\k}}{\partial \Delta^{\prime}}tanh\frac{\beta
E^{-}_{\k}}{2} \right)
\ee
and
\be
\frac{K\Delta^{\prime}+g^{\prime}\Delta}{gg^{\prime}-K^{2}}~=~
\sum_{\k} \left( \frac{\partial E^{+}_{\k}}{\partial \Delta}tanh\frac{\beta
E^{+}_{\k}}{2} +\frac{\partial E^{-}_{\k}}{\partial \Delta}
tanh\frac{\beta E^{-}_{\k}}{2}\right)
\ee

These gap equations are solved numerically and the two gap functions,
have the usual square root dependence on temperature, but opposite signs.
This change of sign of the gap function on the same sheet of the FS has
important bearing on the  physics of the model.  


Since the FS is a combination of contributions from the chain and plane
bands, as we move along any one of sheets of the FS the OP changes from 
$\Delta$ to $\Delta^\prime$ or vice versa, and hence changes sign.
The appearence of OP with different signs can be seen by expanding the 
Free energy, obtained from MF Hamiltonian (Eqn. 6), about
($\Delta, \Delta^{\prime}$) $=$(0,0) close to the transition.The expression 
to second order in $\Delta$ and $\Delta^{\prime}$  is,
\begin{equation}
F=a|\Delta|^{2} + b|\Delta^{\prime}|^{2} + 4c (\Delta\Delta^{\ast} + \Delta^{
\ast}\Delta )
\end{equation}
with $ a=\frac{g^{\prime}}{gg^{\prime}-K^{2}}-u-v-w,
b=\frac{g}{gg^{\prime}-K^{2}}-u-v+w,
 $and$  c=\frac{K}{gg^{\prime}-K^{2}}+v  $\\
where,
$$ u=\frac{1}{4}\sum\limits_{\k}
\left[\frac{tanh \beta E^{+} (0,0)}{{E^+} (0,0)} 
+ \frac{tanh\beta{E^-}(0,0)}{E^{-}(0,0)}\right]  $$ 

$$v=\frac{1}{2}\sum\limits_{\k}
\frac{t^2}{\sqrt{({\eps^2}-{{\eps^\prime}}^2)^{2}+4{t^2}(\eps+{\eps^\prime})^2}}
\left[ \frac{tanh\beta{E^+}(0,0)}{{E^+}(0,0)} 
-\frac{tanh\beta{E^-}(0,0)}{{E^-}(0,0)}\right] $$ 
$$w=\frac{1}{4}\sum\limits_{\k}
\frac{\eps^{2}-{\eps^\prime}^2}{\sqrt{({\eps^2}-{\eps^\prime}^2)^{2}+
4t^2(\eps+{\eps^\prime})^2}} 
\left[\frac{tanh\beta{E^+}(0,0)}{{E^+}(0,0)} -
\frac{tanh\beta{E^-}(0,0)}{{E^-}(0,0)}\right]$$

A nonzero $c$ ensures that there is a single T$_c$, because whenever
 one of the order parameters approaches zero, it forces the other to vanish
simultanously.  Since $v$ can be seen to be positive and the interaction 
strength $K$ is not large enough (${K^2}<g{g^\prime} $), the sign of the  
coefficient of the third term in the expanded free energy above is always 
nonzero (unless $t=K=0$ ) and positive. The free
energy is then minimized  with  the order parameters $\Delta$ and 
$\Delta^{\prime}$ having  opposite signs. This, of course,  ensures 
that there is a node on the FS and that there is a single transition 
temperature.  Experimentally, one does indeed see only one specific 
heat anomaly\cite{leggett}.


Having obtained $\Delta$ and $\Delta^{\prime}$ thus, it is natural to look
for the sign of the OP on the FS. We show in Fig. 3 how $\Delta_{\k}$
behaves as one moves along the (non-interacting) FS. The consequent node
positions on the FS are shown in Fig. 1 for optimal values of $g,
g^{\prime} $ and $K$ (that fit the $T_c$, see discussion below). The nodes
are very close to the $(\pi,\pi)$  
direction as seen experimentally for YBa$_2$Cu$_3$O$_7$, a feature 
shared by the d-wave OP. We also calculate the position of the nodes
as a function of doping (inset of Fig. 1). The node positions turn out to
be not very sensitive to doping $x$ in YBa$_2$Cu$_3$O$_{6+x}$, varying
within $\pm 10^{o}$ about  $45^{o}$ i.e.,  ($\pi,\pi$) direction. Although there
is no experimental data to compare it with presently, this remains a
testable prediction of the model at hand.

It is in the fitness of things that at this point we calculate the 
dependence of the superconducting transition temperature $T_c$ as a
function doping $x$ shown in Fig. 4. We have used as fitting
parameters $g, g^{\prime} $ and $K$, while the band parameters have already 
been fixed to fit the ARPES data as discussed earlier. We chose the values of
these parameters such that $g, g^{\prime}$ and $K$ are within reasonable
limits giving $T_c$ at $x=1$ around 100$^{o}$K. The
slightly higher value ($100^{o}$K as compared to the experimental value
$\simeq 90^{o}$) that we have chosen is to account for the strong 
fluctuations that are present in these systems owing to their intrinsic
low dimensionality\cite{chattop}. A reduction of T$_c$ by about 10 to 20
percent due to fluctuations is generally observed. The optimal values,
thus fixed, turned out to be 225, 140 and 125 $meV$, with $t=50 meV$ fixed
earlier. The $T_c$ is maximum at $x=1$ where the Fermi level is closest to
the van Hove singularity (Fig. 2). As one moves away from this filling,
the DOS at the Fermi level drops drastically and then moves over a
plateau. The $T_c$ versus $x$ curve follows this pattern as expected.
We have also studied the behaviour of $T_c$ as a function of the
interaction parameters $g, g^{\prime} $ and $K$ and the interband hopping
strength $t$.  The variations of $T_c$ with
$g$ and $g^\prime$ show the usual BCS behaviour. Interesting variations
are observed in the dependence of $T_c$ on $K$ and $t$. As the two gap 
parameters $\Delta$ and $\Delta^{\prime}$ have opposite signs, and $t$ 
induces band mixing, $T_c$
goes down quite dramatically with $t$. Conversely, $K$ works counter to
this and so pushes the $T_c$ up showing up as a minimum (that moves
towards lower temperature as $t$ is brought down from 50 to 15$meV$) followed 
by the usual rise as expected in a two band model for
superconductivity\cite{suhl}.  
	
	The two different order parameters in our model are not related 
by any symmetry, except  that they have different signs.  The c-axis 
Josephson current, therefore, does not cancel out completely as in 
the case of a d-wave order parameter, although partial cancellation occurs 
due to the opposite signs of  $\Delta$ and $\Delta^{\prime}$ resulting in a
small (compared to an s-wave OP) but finite current as seen in 
experiments\cite{sun1}.  We found  the node on the FS close to the $(\pi,\pi)$ 
direction, which is not very sensitive to the doping concentration. 
The appearance of the node on the FS ensures a linear temperature 
dependence of penetration depth and {\it low energy excitations} seen in
the thermodynamic measurements. The order parameter in $a-$ ( $b-$ ) 
direction is mainly chain (plane) like and hence  provide a natural
explanation for the observed phase change $\pi$ in SQUID experiments.   

The $s$-wave order parameter is weakly influenced by impurities
(Anderson theorem). In a $d-$wave superconductor, however, scattering between
lobes of different sign on the FS reduces the $T_c$ much faster. This 
averaging out of the sign and consequent sensitivity towards non-magnetic
impurities will occur in the present model as well, albeit at a much
slower rate than in the $d-$wave case. Sign of OP changes in the present
model only by scattering from planes to chains, involving strong 
scattering processes only. Sensitivity of $T_c$ towards non-magnetic impurity 
will, therefore, be intermediate between an $s-$ and a $d-$ wave
superconductor in our case\cite{tobe}, as seen in the high T$_c$ 
materials\cite{leggett,sun2}. The superfluid density, obtained by calculating
the leading order fluctuations over the mean-field state gives an
anisotropic penetration depth\cite{tobe}.

In the $YBCO$ system, the unit cell contains two planes and one chain.
So that a more realistic model would be where these additional features are
present. It is important to note, however, that the conclusions of the
present model 
are not very sensitive to whether we work with a chain and a plane, 
two or more planes or chains, or a combination of them. This a simple model
that incorporates the effect of having more than one OP residing on
different bands that are coupled and shows that with the correct choice
of phase of the different OP can capture many of the unusual features
of the symmetry of OP in these systems. It would be nice to see experimental
results on the position of the node(s) on the FS as a function of doping.
In the $d-$wave case, since the position is dictated by symmetry, there
will be no change as the FS changes shape, while in any of the $s-$wave
models, it will. Though in our model, for a range of doping, the sensitivity
is not too strong.  

In conclusion, we have shown that it is possible to understand many of
the contradictory experimental evidences from a simple and realistic model
applicable to layered superconductors.  


\newpage
\vspace {1.0 cm}
\centerline{\bf FIGURE CAPTIONS}
\vspace {1.0 cm}

\noindent {\bf Fig. 1}
 The two sheets of FS are shown in the first quadrant of 
the BZ.The dotted(solid) line corresponds to $\eps^{-} =0$ ($\eps^{+}=0$).
The filled circles show the position of nodes of OP on FSat $x=0.9$. The 
variation of the node positions on FS with $x$ is shown in the inset.\\
\vspace {.5cm}

\noindent{\bf Fig. 2} 
The DOS for noninteracting quasiparticles is shown. vHS is around 10 meV below 
the Fermi Energy. The inset shows the partialal DOS, where the dotted(solid) 
line corresponds to the  $\eps^{-}$ ($~\eps^{+}~$ ) band.\\
\vspace{0.5 cm}

\noindent{\bf Fig. 3}
The anisotropic OP comes out to be same in two different sheets of FS.In the
figure the dashed (solid) line shows how $\Delta_{k}$($={\Delta_{k}}^{\prime} $
behaves on FS corresponds to $\eps^{-}=0$($\eps^{+} =0$ ) at $x=0.9$. Clearly
two point nodes are visible.\\
\vspace{0.5cm}

\noindent{\bf Fig. 4}
	The behaviour of $T_c$ with doping concentration, shown in this figure
is somewhat similar to the experimental curve.The platue around $x=0.8$ is
prominant and the max. $T_c$ about $x\approx 1 $ is because of the presence of
vHS there.
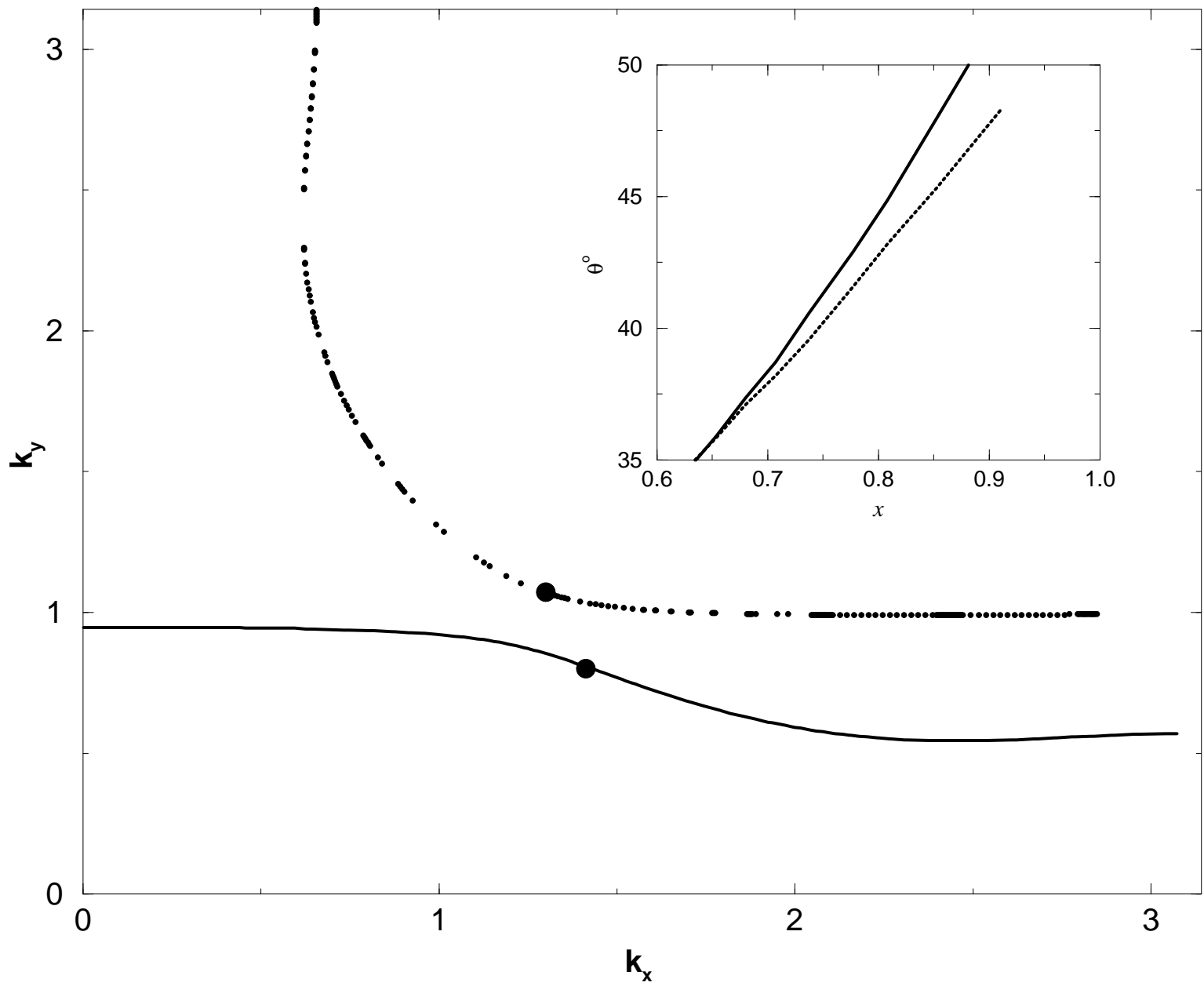
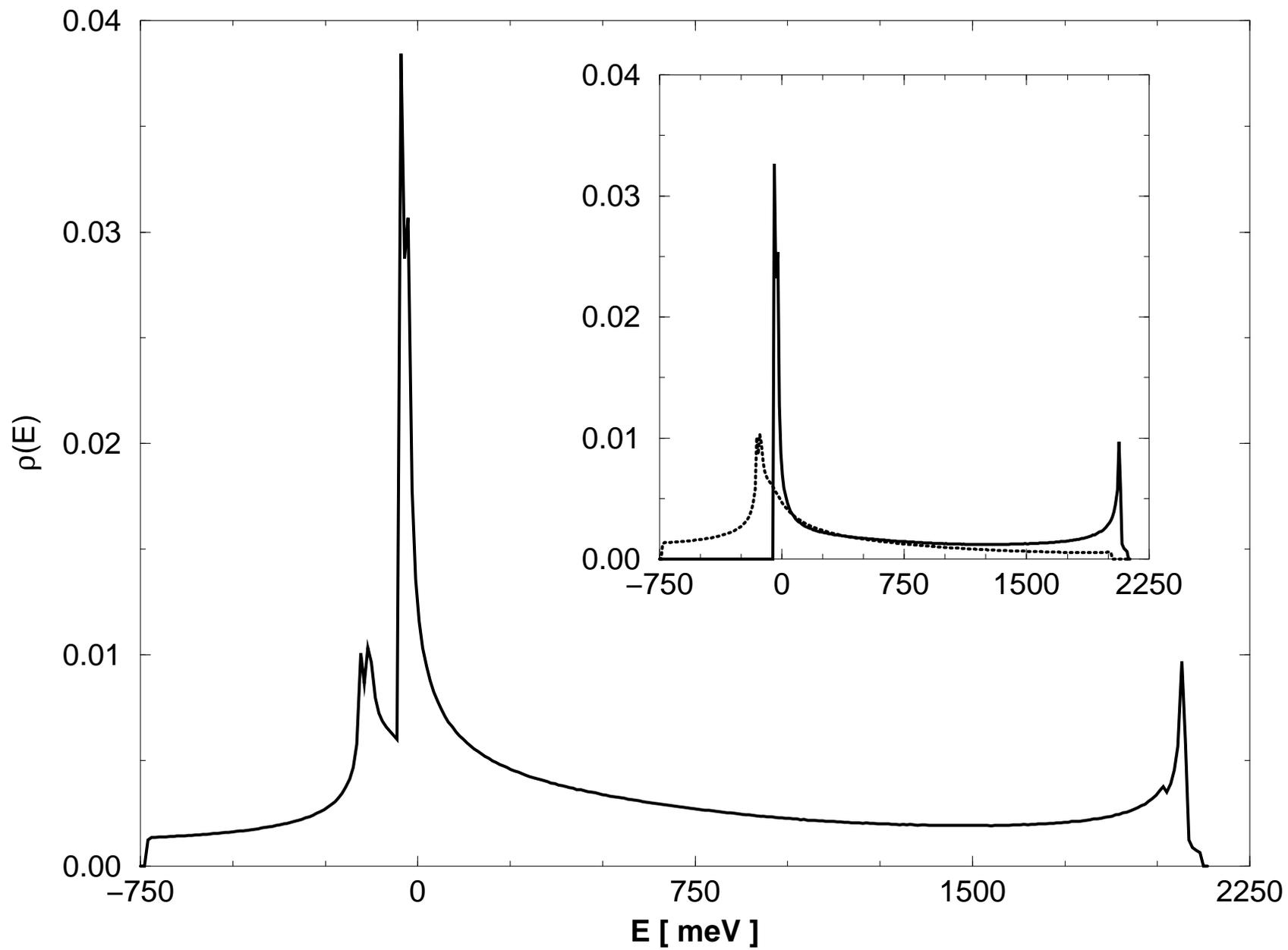
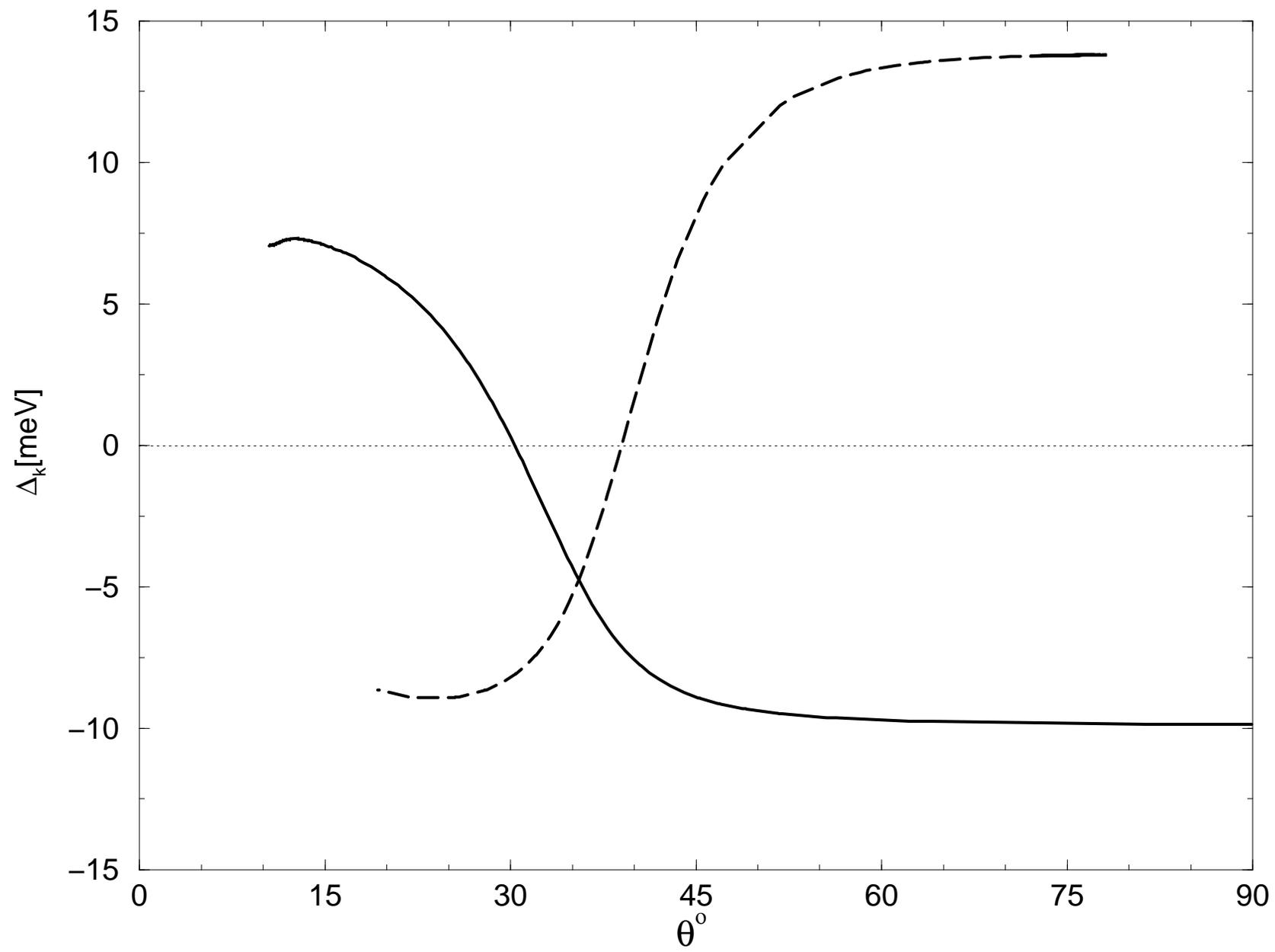
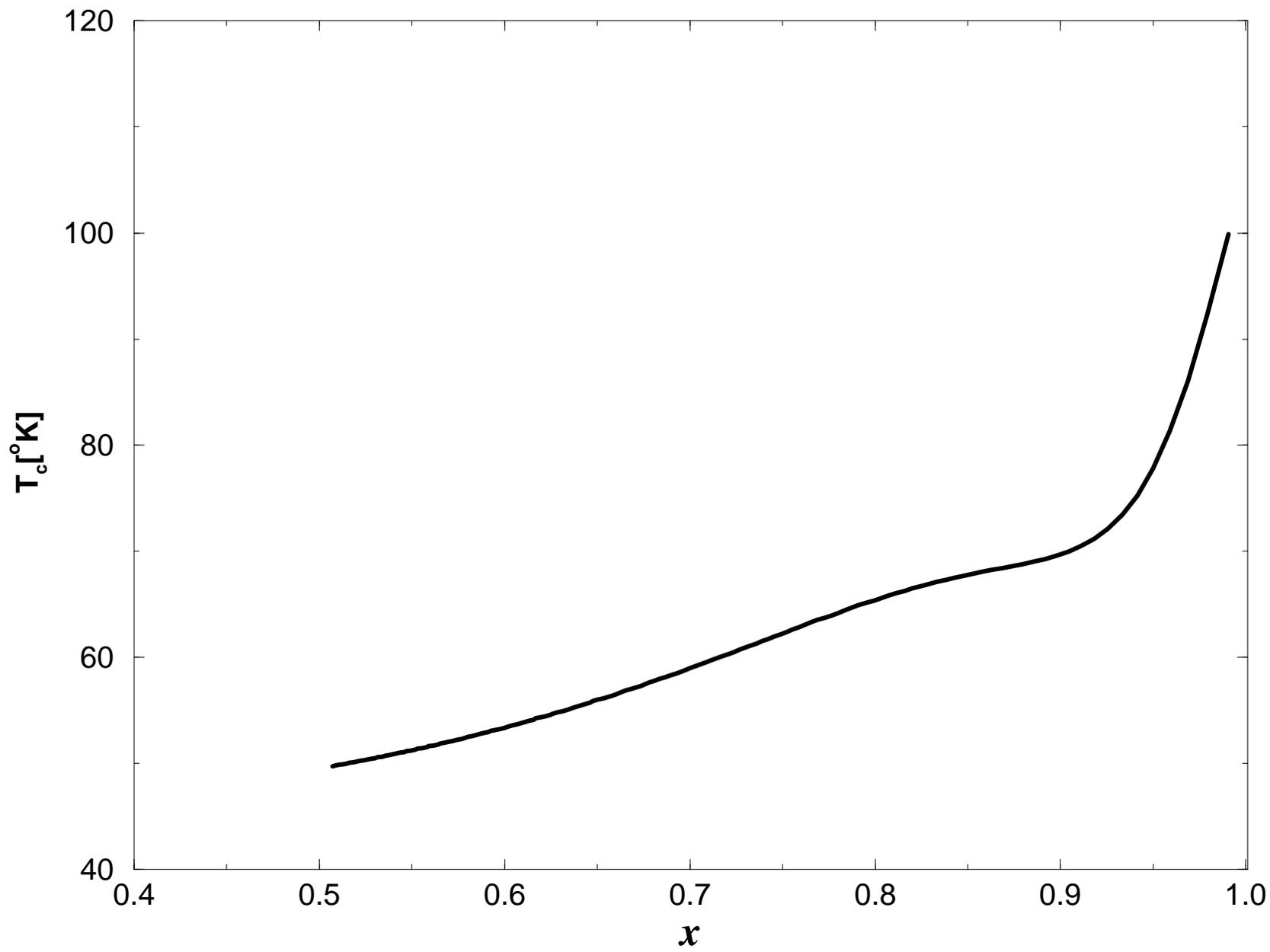
\end{document}